\documentclass[journal,11pt,draftclsnofoot,onecolumn]{IEEEtran}
\usepackage{cite}

\ifCLASSINFOpdf
\else
\usepackage[dvips]{graphicx}
\fi
\usepackage[cmex10]{amsmath}
\usepackage{amsmath,amssymb,bbm,mathrsfs}

\usepackage[T1]{fontenc}
\usepackage{ae}
\usepackage{aecompl}
\usepackage[dvips]{graphicx,color}
\usepackage{bm}
\usepackage{multirow,cite}
\usepackage{type1cm}
\newtheorem{definition}{Definition}[section]
\newtheorem{theorem}{Theorem}[section]
\newtheorem{corollary}{Corollary}[section]
\newtheorem{lemma}{Lemma}[section]

\newtheorem{remark}{Remark}[section]

\IEEEoverridecommandlockouts

\renewcommand{\IEEEQED}{\IEEEQEDopen}

\makeatletter

\@addtoreset{equation}{section}
\makeatother
\makeatletter
    
\@addtoreset{figure}{section}
\makeatother

\begin{document}
%
\title{Optimum Overflow Thresholds in Variable-Length Source Coding Allowing Non-Vanishing Error Probability}
%
%
%

\author{Ryo Nomura,~\IEEEmembership{Member,~IEEE,}
        and~Hideki Yagi,~\IEEEmembership{Member,~IEEE}
\thanks{R. Nomura is with the School of Network and Information, Senshu University, Kanagawa 214-8580, Japan,
 e-mail: nomu@isc.senshu-u.ac.jp}
\thanks{H. Yagi is with the Department of Computer and Network Engineering,  
The University of Electro-Communications, Tokyo 184-8795, Japan, 
e-mail: h.yagi@uec.ac.jp}
\thanks{This paper  is an extended version of the conference paper \cite{NY2017_11} presented at the 2017 IEEE Information Theory Workshop, Kaoshiung, Taiwan.}
}

\maketitle

\begin{abstract}
The variable-length source coding problem allowing the error probability up to some constant is considered for \textit{general} sources.
In this problem the optimum mean codeword length of variable-length codes has already been determined.
On the other hand, in this paper, we focus on the overflow (or excess codeword length) probability instead of the mean codeword length.
The infimum of overflow thresholds under the constraint that both of the  error probability and the overflow probability are smaller than or equal to some constant is called the optimum overflow threshold. 
In this paper, we first derive finite-length upper and lower bounds on these probabilities so as to analyze the optimum overflow thresholds.
Then, by using these bounds we determine the \textit{general} formula of the optimum overflow thresholds in both of the \textit{first-order and second-order} forms.
Next, we consider another expression of the derived {general} formula so as to reveal the relationship with the optimum coding rate in the fixed-length source coding problem.
We also derive \textit{general} formulas of the optimum overflow thresholds in the optimistic coding scenario.
Finally, we apply general formulas derived in this paper to the case of stationary memoryless sources.
\end{abstract}


%
\IEEEpeerreviewmaketitle

\section{Introduction}
The variable-length source coding is one of important problems from both practical and theoretical points of view.
The performance of variable-length codes is evaluated by several criteria such as the mean codeword length, the overflow probability (or excess code length) and so on.
Shannon \cite{Shannon} has first demonstrated that the infimum of the mean codeword length coincides with the source entropy for stationary memoryless sources. 
Han \cite{Han} has extended the results into the case of \textit{general } sources.
The overflow probability, which is defined as the probability of codeword length being above some threshold, has also been analyzed in several contexts \cite{Merhav91:overflow,Uchida01:overflow,KV2014}.
Uchida and Han \cite{Uchida01:overflow} have shown the infimum of achievable thresholds given the overflow probability \textit{exponent} $r$ for \textit{general} sources.
Kontoyiannis and Verd\'{u} \cite{KV2014} have investigated the optimum overflow threshold, which means the infimum of the overflow threshold under the constraint that the overflow probability is smaller than or equal to $\delta > 0$.
All the results mentioned above are for the variable-length coding \textit{without} error.

In this paper, on the other hand, we consider the variable-length coding allowing the error probability up to some constant $\varepsilon>0$, which we call the $\varepsilon$-variable-length coding.
The \textit{first-order} optimum mean codeword length of $\varepsilon$-variable-length codes has been derived by Han \cite{Han}, and  Koga and Yamamoto \cite{KY2005}. 
Kostina, Polyanskiy and Verd\'u \cite{KV2015} have determined the \textit{second-order} optimum mean codeword length of the $\varepsilon$-variable-length codes.
They have revealed that the \textit{second-order} optimum mean codeword length of the $\varepsilon$-variable-length codes has a completely different behavior with that of the variable-length codes \textit{without error}\cite{KV2015}.
Yagi and Nomura \cite{YN2017_7} have also characterized the \textit{first-} and \textit{second-order} optimum mean codeword \textit{cost} of the $\varepsilon$-variable-length codes.

Inspired by the result in \cite{KV2015}, we also focus on the the $\varepsilon$-variable-length coding problem and attempt to investigate the optimum overflow threshold in the problem. 
As we have mentioned above, the \textit{first- and second-order} optimum overflow thresholds in the variable-length coding \textit{without error} have already been studied \cite{KV2014,KS2013}. 
We extend the problem setting to the case of $\varepsilon$-variable-length coding and derive the \textit{general} formula of the \textit{first- and second-order} optimum overflow thresholds.
To this end, we first derive finite-length upper and lower bounds on the error probability and the overflow probability.
Then, using these bounds, we determine the \textit{general} formulas of the \textit{first- and second-order} optimum overflow thresholds.
We also provide another expression of our {general} formulas so as to reveal the relationship with the optimum coding rate in the fixed-length coding problem.

Related works include the work by Saito and Matsushima \cite{saito2016}, in which the \textit{first-order} optimum overflow threshold in $\varepsilon$-variable-length coding has been determined by using the smooth max entropy (or smooth R\'enyi entropy of order zero).
The analyses here are based on information spectrum methods and the approach is different with that in \cite{saito2016}.

This paper is organized as follows. In Section II, we describe the problem setting and give some definitions of the \textit{first-} and \textit{second}-order optimum overflow thresholds.
In Section III, we derive the finite blocklength upper and lower bounds so as to investigate the optimum overflow threshold in the subsequent sections.
In Section IV, we show the \textit{general formula} of the optimum \textit{first-order} overflow threshold. We also give another expression of the \textit{general} formula and compare to the \textit{first}-order optimum achievable rates in the \textit{fixed-length} source coding.
In Section V, we show the \textit{general} formula of the \textit{second}-order optimum thresholds.
Section VI is devoted to the analysis in the optimistic coding scenario. 
In Section VII, we compute the optimum thresholds for the stationary memoryless source by using \textit{general formulas} given in the preceding sections.
Finally, we provide some concluding remarks on our results in Section VIII.
\section{Variable-length coding allowing errors}
\subsection{Problem Setting}
Let ${\bf X} = \left\{X^n = (X_1^{(n)}, X_2^{(n)},\dots,X_n^{(n)})\right\}_{n=1}^\infty$ denote a \textit{general} source, where each $X_i^{(n)}$ takes a value in the finite or countably infinite alphabet ${\cal X}$. 
We use the term \textit{general source} to denote a sequence of random variables $X^n$ indexed by blocklength $n$ and denote the probability distribution of $X^n$ as $P_{X^n}$.
We consider the variable-length codes characterized as follows.
Let
$
\varphi_n : {\cal X}^n \rightarrow {\cal U}^{\ast}$ and $\psi_n :{\cal U}^\ast \rightarrow {\cal X}^n
$
denote a variable-length encoder and a decoder respectively, where
$
{\cal U} = \{ 1,2,\cdots,K\}
$
is a code alphabet and ${\cal U}^{\ast}$ is the set of all finite-length strings over ${\cal U}$ excluding the null string.
The codeword length for the source sequence ${\bf x} \in {\cal X}^n$ is denoted  by $l(\varphi_n({\bf x}))$ when we use the encoder $\varphi_n$.

In this setting, we are interested in the following two probabilities:
\begin{definition}
The error probability of $(\varphi_n,\psi_n)$ and the overflow probability of $(\varphi_n,\psi_n)$ with threshold $\eta_n$ are respectively defined as
\begin{IEEEeqnarray}{rCl}
\varepsilon_n &:= &\Pr \left \{X^n \neq \psi_n(\varphi_n(X^n)) \right \}, \\
\delta_n(\eta_n) & :=& \Pr \left \{l(\varphi_n(X^n)) > \eta_n \right \}.
\end{IEEEeqnarray}
\end{definition}

Notice here that $\eta_n <1$ always leads to $\delta_n(\eta_n)=1$.
Hence, without loss of generality we assume that $\eta_n \ge 1$.
In particular, we consider two cases such as $\eta_n = nR$ and $\eta_n = nR + \sqrt{n}L$.
Let us define the \textit{first}- and \textit{second}-order optimum (overflow) thresholds.
\begin{definition}
Rate $R$ is said to be $(\varepsilon,\delta)$-achievable $(\varepsilon, \delta \in [0,1))$, if there exists a sequence of variable-length code $(\varphi_n,\psi_n)$ such that
\begin{equation} 
\limsup_{n \to \infty} \varepsilon_n \le \varepsilon, \quad
\limsup_{n \to \infty} \delta_n(nR) \le \delta.
\end{equation}
\end{definition}
\begin{definition}[First-order $(\varepsilon,\delta)$-optimum threshold]
\begin{equation}
R(\varepsilon,\delta|{\bf X}) := \inf\left\{ R | R \mbox{ is $(\varepsilon,\delta)$-achievable}  \right\}.
\end{equation}
\end{definition}
The second-order optimum threshold is similarly defined as follows.
\begin{definition}
Rate $L$ is said to be $(\varepsilon,\delta, R)$-achievable $(\varepsilon, \delta \in [0,1), R\ge 0)$, if there exists a sequence of variable-length code $(\varphi_n,\psi_n)$ such that
\begin{equation} 
\limsup_{n \to \infty} \varepsilon_n \le \varepsilon, \quad  
\limsup_{n \to \infty} \delta_n \left(nR + \sqrt{n}{L}\right) \le \delta.
\end{equation}
\end{definition}
\begin{definition}[Second-order $(\varepsilon,\delta,R)$-optimum threshold]
\begin{equation}
L(\varepsilon,\delta, R|{\bf X}) := \inf\{L | L \mbox{ is } (\varepsilon,\delta,R)\mbox{-achievable} \}.
\end{equation}
\end{definition}
\begin{remark}
It is not difficult to check that the condition $\varepsilon + \delta \ge 1$ yields the trivial result such as $R(\varepsilon,\delta|{\bf X}) = 0$ or $L(\varepsilon,\delta, R|{\bf X})= -\infty$. Hence, in this paper we assume that $\varepsilon + \delta < 1$ holds.
\end{remark}

In this paper, we consider the non-prefix variable-length code. 
We here derive the necessary condition for non-prefix variable-length codes allowing errors.
For a variable-length code $(\varphi_n, \psi_n)$,
let $D_n(\varphi_n, \psi_n) \subset {\cal X}^n$ and $T_n(\varphi_n, \eta_n)\subset {\cal X}^n$ be defined as follows: 
\begin{equation}
D_n(\varphi_n, \psi_n) := \left\{ {\bf x} \in {\cal X}^n \left| {\bf x} =  \psi_n(\varphi_n({\bf x})) \right.  \right\},
\end{equation}
\begin{equation}
T_n(\varphi_n, \eta_n) := \left\{ {\bf x}\in {\cal X}^n \left| l(\varphi_n({\bf x})) \le \eta_n \right. \right\}.
\end{equation}
Then, 
since any sequence ${\bf x} \in D_n(\varphi_n, \psi_n)$ is correctly decodable, it holds that
\begin{IEEEeqnarray}{rCl}  \label{eq:condition}
|D_n(\varphi_n, \psi_n) \cap T_n(\varphi_n, \eta_n) | & \le & \sum_{i=1}^{\eta_n} K^i 
<  K^{\eta_n + 1}.
\end{IEEEeqnarray}
We use (\ref{eq:condition}) instead of Kraft's inequality as a condition for non-prefix variable-length codes in this paper.
Throughout this paper, the logarithm is taken to the base $K$.
\subsection{Previous Results}
Saito and Matsushima \cite{saito2016} have derived the \textit{first-order} $(\varepsilon,\delta)$-optimum threshold by using the smooth max entropy.
\begin{definition}[Smooth max entropy]
For any given $\gamma \in [0,1)$, the smooth max entropy (or smooth R\'enyi entropy of order zero) of the source is defined by
\begin{equation} \label{eq:renyi}
H^\gamma(X) := \min_{{ A}\subset {\cal X}: \Pr\{X \in A\} \ge 1-\gamma} \log |A|.
\end{equation}
\end{definition}
\begin{theorem}[Saito and Matsushima \cite{saito2016}]
For any $\varepsilon, \delta \in [0,1)$ satisfying $\varepsilon + \delta <1$, it holds that 
\begin{equation}  \label{eq:renyi2}
R(\varepsilon,\delta|{\bf X}) = \lim_{\nu \downarrow 0} \limsup_{n \to \infty} \frac{1}{n} H^{\varepsilon+\delta+\nu}(X^n).
\end{equation}
\IEEEQED
\end{theorem}
\begin{remark} \label{remark:Uye}
In the fixed-length source coding problem, the infimum of achievable rates under the constraint that the error probability is asymptotically up to $\varepsilon$ is called the $\varepsilon$-optimum coding rate.
Uyematsu \cite{Uyematsu2010} has provided the \textit{general} formula of the $\varepsilon$-optimum coding rate also by using the smooth max entropy.
\end{remark}
\section{Finite Blocklength bounds}
In this section, 
we derive the finite blocklength upper and lower bounds on the error probability and the overflow probability.
\begin{theorem}[Finite blocklength upper bound] \label{theo:direct}
Let $a_n > 0$, $\eta_n \ge 1$ be arbitrary positive numbers. Then, for any $A_n \subset {\cal X}^n$ satisfying $\Pr\{ X^n \in A_n\}\ge 1 - \varepsilon$, there exists a variable-length code $(\varphi_n,\psi_n)$ such that
\begin{IEEEeqnarray}{rCl}  \label{eq:direct1}
\varepsilon_n & \le & \varepsilon, \nonumber \\
\delta_n(\eta_n)  & \le & \Pr\left\{  a_n \frac{P_{X^n}({X^n})}{ \Pr \left\{ X^n \in A_n \right  \}} \le K^{-\eta_n}, \  X^n \in A_n \right\}  + a_nK.
\end{IEEEeqnarray}
\end{theorem}
\begin{IEEEproof}
We first construct the encoder and the decoder.

[Encoder $\varphi_n$]: 
For any fixed $A_n \subset {\cal X}^n$ satisfying $\Pr\{ X^n \in A_n\}\ge 1 - \varepsilon$, we define the encoder as
\begin{equation}
\varphi_n({\bf x}) = \left\{ \begin{array}{cc}
f_n({\bf x})  & {\bf x} \in A_n, \\
1 & \mbox{otherwise},
 \end{array}   \right.
\end{equation}
where $f_n: A_n \to {\cal U}^\ast$ is an injection mapping which assigns the codeword whose length is $\lceil -\log \frac{P_{X^n}({\bf x})}{\Pr \{ X^n \in A_n\} } \rceil $ to each ${\bf x} \in A_n$.
It is not difficult to verify that there exists such an injection mapping.
Then, the codeword length $l(\varphi_n({\bf x}) )$ of this code is given by
\begin{equation}
l(\varphi_n({\bf x}) )= \left\{ \begin{array}{cc}
\lceil -\log \frac{P_{X^n}({\bf x})}{\Pr \{ X^n \in A_n\} } \rceil  & {\bf x} \in A_n, \\
1 & \mbox{otherwise}.
 \end{array}   \right.
\end{equation}

[Decoder $\psi_n$]: 
The decoder $\psi_n$ is arranged to be an inverse mapping of $f_n$.
That is, for a received sequence ${\bf u} \in {\cal U}^\ast$, if there exists ${\bf x} \in A_n$ such that  ${\bf u} = f_n({\bf x})$, then the decoder outputs $\psi_n({\bf u}) = {\bf x}$.
If there does not exist such ${\bf x}$ (this would happen, for example, when $f_n({\bf x}) \neq 1$ holds for all ${\bf x} \in A_n$), then the decoder declares an error.

Next, we evaluate the error probability and the overflow probability of this variable-length code.
From the construction of the code, the error probability is given by
$
\varepsilon_n \le \Pr\left\{ X^n \notin A_n \right\} \le \varepsilon.
$
Hence, it suffices to show (\ref{eq:direct1}). To do so, let us define $S_n$ and $B_n$ as follows:
\begin{align} \label{eq:sn}
S_n & := \left\{ {\bf x} \in {\cal X}^n \left| l(\varphi_n({\bf x})) > \eta_n \right. \right\}, \qquad  \\
B_n & :=  \left\{ {\bf x} \in {\cal X}^n \left| a_n \frac{P_{X^n}({\bf x})}{ \Pr \left\{ X^n \in A_n \right\}} \le K^{-\eta_n }  \right. \right\}. 
\end{align}
Then, since $\eta_n \ge 1$ holds, we have
\begin{equation} \label{eq:1-0}
(A_n)^c \subseteq (S_n)^c
\end{equation}
from the construction of the code $(\varphi_n,\psi_n)$, where $c$ denotes the complement. Moreover, for any ${\bf x} \in S_n$ it holds that
$
-\log \frac{P_{X^n}({\bf x})}{\Pr \{ X^n \in A_n\} } > \eta_n - 1.
$
This means that for any ${\bf x} \in S_n$
\begin{equation} \label{eq:1-1}
P_{X^n}({\bf x}) < K^{ -( \eta_n - 1)}{\Pr \{ X^n \in A_n\} }.
\end{equation}
Thus, from (\ref{eq:1-0}) and (\ref{eq:1-1}) we have
\begin{IEEEeqnarray}{rCl}  \label{eq:1-2}
\delta_n(\eta_n)
& = &  \Pr \left\{ X^n \in S_n \cap B_n \right\} +  \Pr \left\{ X^n \in S_n \cap (B_n)^c \right\} \nonumber \\
& \le & \Pr \left\{ X^n \in A_n \cap B_n \right\} +  \sum_{{\bf x} \in S_n \cap (B_n)^c} P_{X^n}({\bf x}) \nonumber \\
& \le & \Pr \left\{ X^n \in A_n \cap B_n \right\} +  \sum_{{\bf x} \in (B_n)^c} K^{ -( \eta_n - 1)}{\Pr \{ X^n \in A_n\} } \nonumber \\
& \le & \Pr \left\{ X^n \in A_n \cap B_n \right\} +  |(B_n)^c | K^{ -( \eta_n - 1)}{\Pr \{X^n \in  A_n\} }.
\end{IEEEeqnarray}
Next, we evaluate the second term on the r.h.s. of (\ref{eq:1-2}).
From the definition of $B_n$, we have
\begin{IEEEeqnarray}{rCl}
1 & \ge & \sum_{{\bf x} \notin B_n } P_{X^n}({\bf x}) \nonumber \\ 
&\ge & \sum_{{\bf x} \notin B_n} \frac{K^{-\eta_n}}{a_n} \Pr\left\{X^n \in  A_n \right\} \nonumber \\
& = & |(B_n)^c|  \frac{K^{-\eta_n}}{a_n} \Pr\left\{ X^n \in A_n \right\}
\end{IEEEeqnarray}
from which it follows that
\begin{IEEEeqnarray}{rCl} \label{eq:1-1-1}
|(B_n)^c| \le a_n K^{\eta_n} \frac{1}{\Pr\left\{ X^n \in A_n \right\}}.
\end{IEEEeqnarray}
Plugging (\ref{eq:1-1-1}) into (\ref{eq:1-2}) yields (\ref{eq:direct1}).
\end{IEEEproof}
%
%
%
%
\begin{theorem}[Finite blocklength lower bound] \label{theo:converse}
For an arbitrary fixed variable-length code $(\varphi_n,\psi_n)$, we set 
$D_n = \left\{ {\bf x} \in {\cal X}^n \left| {\bf x} = \psi_n(\varphi_n({\bf x})) \right. \right\}.$
Then, for any $a_n>0$ and any $\eta_n \ge 1$ it holds that
\begin{IEEEeqnarray}{rCl} 
\delta_n(\eta_n)  
 \ge & \Pr\left\{  \frac{P_{X^n}({X^n})}{ \Pr \left\{ X^n \in D_n \right  \}} \le a_n K^{-\eta_n}, \  X^n \in D_n \right\}  - a_nK\Pr\left\{ X^n \in D_n \right\}.
\end{IEEEeqnarray}
\end{theorem}
\begin{IEEEproof}
Set $\tilde{B}_n$ as
\begin{align}
\tilde{B}_n & :=  \left\{ {\bf x} \in {\cal X}^n \left| \frac{P_{X^n}({\bf x})}{ \Pr \left\{ X^n \in D_n \right\}} \le a_n K^{-\eta_n }  \right. \right\} 
\end{align}
and $S_n$ as in (\ref{eq:sn}).
Then, we have
\begin{IEEEeqnarray}{rCl}  \label{eq:converse1}
\lefteqn{ \Pr\left\{  \frac{P_{X^n}({X^n})}{ \Pr \left\{ X^n \in D_n \right\}} \le a_n K^{-\eta_n} , \   X^n \in D_n  \right\}} \nonumber \\
 & = &  \sum_{{\bf x} \in D_n \cap \tilde{B}_n \cap S_n} P_{X^n}({\bf x}) +  \sum_{{\bf x} \in D_n \cap \tilde{B}_n \cap (S_n)^c} P_{X^n}({\bf x}) \nonumber \\
& \le & \delta_n(\eta_n) + \sum_{{\bf x} \in D_n \cap (S_n)^c} a_nK^{-\eta_n} \Pr\left\{ X^n \in D_n \right\} \nonumber \\
& = & \delta_n(\eta_n) + \left| D_n \cap (S_n)^c \right|  a_nK^{-\eta_n} \Pr\left\{ X^n \in D_n \right\} \nonumber \\
& \le & \delta_n(\eta_n) +  a_nK \Pr\left\{ X^n \in D_n \right\},
\end{IEEEeqnarray}
where the last inequality is due to (\ref{eq:condition}).
This completes the proof of the theorem.
\end{IEEEproof}
\section{First-order $(\varepsilon,\delta)$-optimum threshold}
\subsection{General Formula}
In this section, we establish the general formula of the first-order $(\varepsilon,\delta)$-optimum threshold by using Theorems \ref{theo:direct} and \ref{theo:converse}.
We define the quantity ${G}_{\varepsilon,\delta}({\bf X})$ as
\begin{IEEEeqnarray}{rCl}  \label{eq:generalformula1}
\lefteqn{{G}_{\varepsilon,\delta}({\bf X})} \nonumber \\
&  := & \inf\left\{  R \left| \lim_{\nu \downarrow 0}\limsup_{n \to \infty}\inf_{ \substack{A_n \subset {\cal X}^n:\\ \Pr\{X^n \in A_n\} \ge 1- \varepsilon - \nu}} \right. \Pr\left\{ \!- \frac{1}{n}\log \frac{P_{X^n}(X^n)}{\Pr\{X^n \!\in\! A_n\}} \!\ge\! R, X^n \!\in\! A_n   \right\} \le \delta  \right\}.
\end{IEEEeqnarray}
Then, we have the following theorem:
\begin{theorem}[First-order $(\varepsilon,\delta)$-optimum threshold] \label{theo:general1}
For any $\varepsilon, \delta \in [0,1)$ satisfying $\varepsilon + \delta < 1$, it holds that
\begin{align}
R(\varepsilon,\delta|{\bf X}) & = {G}_{\varepsilon,\delta}({\bf X}).
\end{align}
\end{theorem}
\begin{IEEEproof} The proof consists of two parts.

(Direct Part:)
Setting $R_0$ as 
\begin{IEEEeqnarray}{rCl}
R_0 := {G}_{\varepsilon,\delta}({\bf X}),
\end{IEEEeqnarray}
we show that for any $\gamma > 0$, $R=R_0 + 2\gamma$ is $(\varepsilon,\delta)$-achievable.
To do so, we arbitrarily fix $\nu \in (0, 1-\varepsilon]$ and use Theorem \ref{theo:direct} with 
$a_n = K^{-n \gamma}$ and $\eta_n= nR=n(R_0 + 2\gamma)$.
Let $\lambda_1 > \lambda_2 > \dots \to 0$ be an arbitrary decreasing sequence.
We choose $A_n \subseteq {\cal X}^n$ satisfying 
$\Pr\left\{  X^n \in A_n \right\} \ge 1 - \varepsilon - \nu$
and 
\begin{IEEEeqnarray}{rCl} \label{eq:d-2-2-2}
\lefteqn{
\Pr\left\{  -\frac{1}{n} \log \frac{P_{X^n}({X^n})}{ \Pr \left\{ X^n \in A_n \right  \}} \ge R_0 + \gamma , \  X^n \in A_n \right\}
} \nonumber  \\
 & \le &\inf_{\substack{A_n \subset {\cal X}^n:\\ \Pr\{X^n \in A_n\} \ge 1- \varepsilon - \nu}} \Pr\left\{  -\frac{1}{n} \log \frac{P_{X^n}({X^n})}{ \Pr \left\{ X^n \in A_n \right  \}} \ge R_0 \!+\! \gamma , \  X^n \in A_n \right\} \!+\! \lambda_n.
 \end{IEEEeqnarray}
Then, for this $A_n \subseteq {\cal X}^n$, from Theorem \ref{theo:direct} there exists a variable-length code $(\varphi_n,\psi_n)$ such that 
\begin{equation} \label{eq:2-2-2-3}
\varepsilon_n \le \varepsilon + \nu,
\end{equation}
and
\begin{IEEEeqnarray}{rCl} \label{eq:2-2-2-4}
\delta_n(nR) & \!=\! & \Pr \left\{ \frac{1}{n} l(\varphi_n(X^n)) > R_0 + 2\gamma \right\} \nonumber \\
& \!\le\! & \Pr\left\{ \! -\frac{1}{n} \log \frac{P_{X^n}({X^n})}{ \Pr \left\{ X^n \in A_n \right  \}} \ge R_0 \!+\! \gamma , \  X^n \in A_n \right\} + K^{-n\gamma +1}.
\end{IEEEeqnarray}
It follows from (\ref{eq:d-2-2-2}) that 
\begin{IEEEeqnarray}{rCl}  \label{eq:d-2-2-2-4}
\lefteqn{\limsup_{n \to \infty} \Pr \left\{ \frac{1}{n} l(\varphi_n(X^n)) > R_0 + 2\gamma \right\}} \nonumber  \\
 & \le &\limsup_{n \to \infty}  \inf_{\substack{A_n \subset {\cal X}^n:\\ \Pr\{X^n \in A_n\} \ge 1- \varepsilon - \nu}}  \Pr\left\{  -\frac{1}{n} \log \frac{P_{X^n}({X^n})}{ \Pr \left\{ X^n \in A_n \right  \}} \ge R_0 + \gamma , \  X^n \in A_n \right\} \nonumber \\
& \le & \delta,
\end{IEEEeqnarray}
where the last inequality is due to the definition of $R_0$. From (\ref{eq:2-2-2-3}), (\ref{eq:2-2-2-4}) and  (\ref{eq:d-2-2-2-4}), the direct part has been proved.

(Converse Part:)
We assume that $R$ is $(\varepsilon,\delta)$-achievable. Then, it holds that
\begin{align}  \label{eq:c-1}
\limsup_{n \to \infty} \varepsilon_n &\le \varepsilon, \\ \label{eq:c-2}
\limsup_{n \to \infty} \delta_n(nR) &\le \delta.
\end{align}
By using Theorem \ref{theo:converse} with $a_n = K^{-n \gamma} \ (\forall \gamma> 0)$ and $\eta_n= nR$, we have
\begin{IEEEeqnarray}{rCl} 
\delta_n(nR) & \ge & \Pr\left\{  \frac{P_{X^n}({X^n})}{ \Pr \left\{ X^n \!\in\! D_n \right  \}} \le K^{-n(R + \gamma)}, \  X^n \!\in\! D_n \right\} - K^{-n\gamma +1},
\end{IEEEeqnarray}
where 
$
D_n = \left\{ {\bf x} \in {\cal X}^n \left| {\bf x} = \psi_n(\varphi_n({\bf x})) \right. \right\}.
$
Here, (\ref{eq:c-1}) means that
$ \Pr\left\{ X^n \in D_n \right\} \ge 1 - \varepsilon - \nu $ \ \ $(\forall n > n_0)$ holds  for any $\nu \in (0, 1-\varepsilon)$.
Thus, for any $\nu \in (0, 1-\varepsilon)$ we have
\begin{IEEEeqnarray}{rCl} 
\lefteqn{\limsup_{n \to \infty}\delta_n(nR)} \nonumber \\
 & \ge & \limsup_{n \to \infty} \Pr\left\{ \!-\frac{1}{n} \log \frac{P_{X^n}({X^n})}{ \Pr \left\{ X^n \in D_n \right  \}} \ge R \!+\! \gamma , \  X^n \!\in\! D_n \right\} \nonumber \\
 & \ge & \limsup_{n \to \infty} \inf_{\substack{A_n \subset {\cal X}^n:\\ \Pr\{X^n \in A_n\} \ge 1- \varepsilon - \nu}} \Pr\left\{ -\frac{1}{n} \log \frac{P_{X^n}({X^n})}{ \Pr \left\{ X^n \in A_n \right  \}} \ge R + \gamma , \  X^n \in A_n \right\}.
\end{IEEEeqnarray}
Substituting this inequality into (\ref{eq:c-2}), we obtain
\begin{IEEEeqnarray}{rCl} 
\delta 
& \ge & \limsup_{n \to \infty} \inf_{\substack{A_n \subset {\cal X}^n:\\ \Pr\{X^n \in A_n\} \ge 1- \varepsilon - \nu}} \Pr\left\{ \!-\frac{1}{n} \log \frac{P_{X^n}({X^n})}{ \Pr \left\{ X^n \!\in\! A_n \right  \}} \ge R \!+\! \gamma , \  X^n \!\in\! A_n \right\}.
 \end{IEEEeqnarray}
This means that
\begin{IEEEeqnarray}{rCl}
R + \gamma& \ge \inf\left\{  R \left| \lim_{\nu \downarrow 0}\limsup_{n \to \infty}\inf_{\substack{A_n \subset {\cal X}^n:\\ \Pr\{X^n \in A_n\} \ge 1- \varepsilon - \nu}} \right.  \Pr\left\{ - \frac{1}{n}\log \frac{P_{X^n}(X^n)}{\Pr\{X^n \in A_n\}} \ge R, X^n \in A_n   \right\} \le \delta  \right\}, \nonumber \\
\end{IEEEeqnarray}
which implies that the converse part holds.
\end{IEEEproof}
%
%
%
\subsection{Another Expression of the General Formula}
The \textit{general formula} of the $(\varepsilon,\delta)$-optimum thresholds derived in the previous subsection seems complicated and hard to compute even for tractable sources such as stationary memoryless sources, Markov sources and so on.
Hence in this subsection, we derive another expression of ${G}_{\varepsilon,\delta}({\bf X})$ which enables us not only to compute the $(\varepsilon,\delta)$-optimum thresholds for tractable sources but also to understand the structure of the optimum overflow thresholds in the $\varepsilon$-variable-length coding.
As a result, the relationship with the $\gamma$-optimum coding rate in the fixed-length coding is revealed. 

Set
\begin{IEEEeqnarray}{rCl} \label{eq:F}
{F}(\varepsilon,R)  := \limsup_{n \to \infty}\inf_{\substack{A_n \subset {\cal X}^n:\\ \Pr\{X^n \in A_n\} \ge 1- \varepsilon}} 
  \Pr \left\{ \!- \frac{1}{n}\log P_{X^n}(X^n)\!\ge\! R,\ X^n \!\in\! A_n  \right \}.
\end{IEEEeqnarray}
Then, the function ${F}(\varepsilon,R)$ is a monotonically nonincreasing function of $\varepsilon$ and $R$.
It is not difficult to derive the following lemma:
\begin{lemma}  \label{coro:4-1}
\begin{align}  \label{eq:G}
{G}_{\varepsilon,\delta}({\bf X}) & = \inf\left\{  R \left| \lim_{\nu \downarrow 0} F(\varepsilon + \nu,R)\le \delta \right. \right\}.
\end{align}\IEEEQED
\end{lemma}
Here, we define two quantities
\begin{IEEEeqnarray}{rCl} 
\tilde{H}_{\varepsilon,\delta}({\bf X})  := \inf \left\{R \left| {F}(\varepsilon,R) \le \delta \right. \right\},
\end{IEEEeqnarray}
and
\begin{IEEEeqnarray}{rCl}
\overline{H}_{\gamma}({\bf X}) \!:=\! \inf\left\{ R \left| \limsup_{n \to \infty} \Pr \left\{ \frac{1}{n}\log \frac{1}{P_{X^n}(X^n)} \!>\!R  \right\} \!\le\! \gamma \right. \right\}.
\end{IEEEeqnarray}
It should be noted that the $\gamma$-optimum coding rate in the fixed length coding is characterized by $\overline{H}_{\gamma}({\bf X})$ \cite{Han} (cf. Remark \ref{remark:4-1} below).

Then, the following theorem holds:
\begin{theorem} \label{theo:another_ex}
For any $\varepsilon, \delta \in [0,1)$ satisfying $\varepsilon + \delta < 1$, it holds that
\begin{equation}
{G}_{\varepsilon,\delta}({\bf X}) = \tilde{H}_{\varepsilon,\delta}({\bf X})= \overline{H}_{\varepsilon+\delta}({\bf X}) .
\end{equation} \IEEEQED
\end{theorem}
Since $\overline{H}_{\gamma}({\bf X})$ is a right-continuous function of $\gamma$ (see, \cite{Han}), the function ${G}_{\varepsilon,\delta}({\bf X})$ and $\tilde{H}_{\varepsilon,\delta}({\bf X})$ are also right-continuous functions of $\varepsilon$ and $\delta$.

From the theorem we immediately have:
\begin{corollary} \label{coro:2}
Fix $\varepsilon \in [0,1)$ arbitrarily. Then, for any $\varepsilon_1, \varepsilon_2 \ge 0$ satisfying $\varepsilon_1+\varepsilon_2=\varepsilon$ it holds that
\[
\tilde{H}_{\varepsilon_1,\varepsilon_2}({\bf X})= \overline{H}_{\varepsilon}({\bf X}).
\]\IEEEQED
\end{corollary}
\begin{remark} \label{remark:4-1}
The $\varepsilon$-optimum coding rate in the fixed-length coding and the (first-order) $\varepsilon$-optimum threshold in the variable-length coding \textit{without error} coincide with $R(\varepsilon,0|{\bf X})$ and $R(0,\varepsilon|{\bf X})$, respectively. 
Then, the following relation has already been shown in \cite{Hayashi,Nomura2011}．
\begin{IEEEeqnarray}{rCl} \label{eq:4-1-1}
R(\varepsilon,0|{\bf X}) = R(0,\varepsilon|{\bf X}) = \overline{H}_\varepsilon({\bf X}).
\end{IEEEeqnarray}
This equality reveals a deep relationship between the fixed-length coding and the variable-length coding \textit{without error}.
From the above equality and Corollary \ref{coro:2}, for any $\varepsilon_1, \varepsilon_2 \ge 0$ satisfying $\varepsilon_1+\varepsilon_2=\varepsilon$, we obtain the following relation
\begin{IEEEeqnarray}{rCl} \label{eq:4-1-2}
R(\varepsilon_1,\varepsilon_2|{\bf X}) = R(\varepsilon,0|{\bf X}) = R(0,\varepsilon|{\bf X}). 
\end{IEEEeqnarray}
It should be emphasized that the relation (\ref{eq:4-1-2}) subsumes (\ref{eq:4-1-1}),  because $\varepsilon_1$ and $\varepsilon_2$  in (\ref{eq:4-1-2}) may be arbitrary nonnegative numbers satisfying $\varepsilon_1+\varepsilon_2=\varepsilon$.
This relation can also be obtained from the fact that the r.h.s. of (\ref{eq:renyi2}) coincides with $\overline{H}_{\varepsilon+\delta}({\bf X})$ \cite{Uyematsu2010}.
\end{remark}

Before proving Theorem \ref{theo:another_ex}, we show the following lemma.
\begin{lemma} \label{lemma:4-1}
For any $\nu >0$, it holds that
\begin{align} 
\tilde{H}_{\varepsilon+\nu,\delta}({\bf X}) \le {G}_{\varepsilon,\delta}({\bf X}) = \lim_{\nu \downarrow 0} \tilde{H}_{\varepsilon+\nu,\delta}({\bf X}) \le \tilde{H}_{\varepsilon,\delta}({\bf X}).
\end{align}
\end{lemma}
\begin{IEEEproof}[Proof of Lemma \ref{lemma:4-1}]
Since $\tilde{H}_{\varepsilon,\delta}({\bf X})$ is a monotonically nonincreasing function, 
the first and second inequalities clearly hold.
Hence, it suffices to show the intermediate equality.

Since $F(\varepsilon,R)$ is a monotonically nonincreasing function of $\varepsilon$ and $R$, ${G}_{\varepsilon,\delta}({\bf X}) $ can be expressed as
\begin{IEEEeqnarray}{rCl}  \label{eq:4-2-1}
{G}_{\varepsilon,\delta}({\bf X}) & = & \inf \bigcap_{\nu > 0}\left\{  R \left| F(\varepsilon\!+\!\nu,R)\!\le\! \delta \right. \right\}.
\end{IEEEeqnarray}
For any $\nu >0$, we have
\begin{IEEEeqnarray}{rCl} 
{G}_{\varepsilon,\delta}({\bf X}) & \ge & \inf \left\{  R \left| F(\varepsilon + \nu,R) \le  \delta \right. \right\} = \tilde{H}_{\varepsilon+\nu,\delta}({\bf X}),
\end{IEEEeqnarray}
which implies 
\begin{IEEEeqnarray}{rCl}  \label{eq:4-2-2}
{G}_{\varepsilon,\delta}({\bf X}) & \ge & \sup_{\nu > 0}\tilde{H}_{\varepsilon+\nu,\delta}({\bf X}) \nonumber \\
& = & \lim_{\nu \downarrow 0}\tilde{H}_{\varepsilon+\nu,\delta}({\bf X}). 
\end{IEEEeqnarray}
On the otherhand, again for any $\nu >0$, we have
\begin{IEEEeqnarray}{rCl}
\lim_{\nu \downarrow 0}\tilde{H}_{\varepsilon+\nu,\delta}({\bf X}) & \ge & \tilde{H}_{\varepsilon+\nu,\delta}({\bf X}) \nonumber \\
& = & \inf\left\{R | F(\varepsilon + \nu, R) \le \delta \right\}.
\end{IEEEeqnarray}
This inequality for an arbitrary fixed $\nu >0$ implies that 
\begin{IEEEeqnarray}{rCl}  \label{eq:4-2-3}
\lim_{\nu \downarrow 0}\tilde{H}_{\varepsilon+\nu,\delta}({\bf X}) & \ge & \inf \bigcap_{\nu >0} \left\{R | F(\varepsilon + \nu, R) \le \delta \right\} \nonumber \\
& = & {G}_{\varepsilon,\delta}({\bf X}),
\end{IEEEeqnarray}
where the equality is due to (\ref{eq:4-2-1}).
Combining (\ref{eq:4-2-2}) and (\ref{eq:4-2-3}) yields
\begin{equation} 
{G}_{\varepsilon,\delta}({\bf X}) = \lim_{\nu \downarrow 0} \tilde{H}_{\varepsilon+\nu,\delta}({\bf X}).
\end{equation}
This completes the proof of the lemma.
\end{IEEEproof}

\begin{IEEEproof}[Proof of Theorem \ref{theo:another_ex}]
From Lemma \ref{lemma:4-1}, it suffices to prove two inequalities:
\begin{IEEEeqnarray}{rCl} \label{eq:4-3}
{G}_{\varepsilon,\delta}({\bf X}) & \ge \overline{H}_{\varepsilon+\delta}({\bf X}), \\ \label{eq:4-4}
\tilde{H}_{\varepsilon,\delta}({\bf X}) & \le \overline{H}_{\varepsilon+\delta}({\bf X}).
\end{IEEEeqnarray}

{(Proof of (\ref{eq:4-3}):)}
For any fixed $R>{G}_{\varepsilon,\delta}({\bf X})$, we show that $R\ge \overline{H}_{\varepsilon+\delta}({\bf X})$ holds.

Set $J_n \subseteq {\cal X}^n$ as 
\begin{equation} \label{eq:bn}
J_n := \left\{{\bf x} \in {\cal X}^n \left| - \frac{1}{n}\log P_{X^n}({\bf x}) \ge R \right. \right\}.
\end{equation}
Then,  by the assumption $R>{G}_{\varepsilon,\delta}({\bf X})$ and (\ref{eq:G}), it holds that
\begin{align}  \label{eq:4-3-1}
\delta & \ge \lim_{\nu \downarrow 0}\limsup_{n \to \infty}\inf_{\substack{A_n \subset {\cal X}^n:\\ \Pr\{X^n \in A_n\} \ge 1- \varepsilon - \nu}}  \Pr\left\{ X^n \in A_n \cap J_n  \right\} \nonumber \\
& \ge \limsup_{n \to \infty}\inf_{\substack{A_n \subset {\cal X}^n:\\ \Pr\{X^n \in A_n\} \ge 1- \varepsilon - \nu}}  \Pr\left\{ X^n \in A_n \cap J_n  \right\}
\end{align}
for any $\nu \in (0,1-\varepsilon)$. Moreover, we define a subset $K_n \subseteq {\cal X}^n$ such that
\begin{IEEEeqnarray}{rCl} \label{eq:4-3-2}
\Pr\left\{ X^n \in K_n \right\} \ge 1 - \varepsilon - \nu
\end{IEEEeqnarray}
and 
\begin{IEEEeqnarray}{rCl}
\Pr\left\{ X^n \in J_n \!\cap\! K_n \right\}
& \le & \inf_{\substack{A_n \subset {\cal X}^n:\\ \Pr\{X^n \in A_n\} \ge 1- \varepsilon - \nu}}  \Pr\left\{ X^n \!\in\! A_n \cap J_n  \right\} + \nu
\end{IEEEeqnarray}
hold. Then, from (\ref{eq:4-3-1}) it holds that
\begin{align}
\limsup_{n \to \infty} \Pr \left\{ X^n \in J_n \cap K_n \right\} \le \delta + \nu.
\end{align}
On the other hand, from (\ref{eq:4-3-2}), we have
\begin{align} 
\Pr \left\{ X^n \in J_n \right\} & \le \Pr \left\{ X^n \in J_n \cap K_n \right\} +  \Pr \left\{ X^n \in (K_n)^c\right\}  \nonumber \\
& \le \Pr \left\{ X^n \in J_n \cap K_n \right\} + \varepsilon + \nu.
\end{align}
This means that
\begin{align} \label{eq:4-8}
\limsup_{n \to \infty}\Pr \left\{ X^n \!\in\! J_n \right\}& \le \limsup_{n \to\infty}\Pr \left\{ X^n \in J_n \cap K_n \right\} + \varepsilon + \nu \nonumber \\
& \le  \delta + \varepsilon + 2\nu.
\end{align}
Since $\nu \in (0,1-\varepsilon)$ is arbitrarily, we have
\begin{align} \label{eq:4-9}
\limsup_{n \to \infty}\Pr \left\{ X^n \in J_n \right\} & \le \delta + \varepsilon,
\end{align}
which implies $R \ge \overline{H}_{\varepsilon+\delta}({\bf X})$.

{(Proof of (\ref{eq:4-4}):)}
For any fixed $R>\overline{H}_{\varepsilon+\delta}({\bf X})$, we show that $R\ge \tilde{H}_{\varepsilon,\delta}({\bf X})$ holds．
We also use the set $J_n$ defined in (\ref{eq:bn}). Since $R>\overline{H}_{\varepsilon+\delta}({\bf X})$ holds, we have
\begin{align} \label{eq:4-10}
\limsup_{n \to \infty} \Pr\left\{ X^n \in J_n \right\} \le \varepsilon + \delta.
\end{align}
Here, without loss of generality we assume that the elements of ${\cal X}^n$ be ordered $\{{\bf x}_1,{\bf x}_2,{\bf x}_3,\cdots \}$ as its probability, that is, for any $i <j$, $P_{X^n}({\bf x}_i) \ge P_{X^n}({\bf x}_j)$ holds.
We define a positive integer $i^\ast$ and $L_n=\{{\bf x}_1,{\bf x}_2,\cdots,{\bf x}_{i^\ast}\}$ such that
\begin{IEEEeqnarray}{rCl} \label{eq:4-11}
\Pr\left\{X^n \in L_n \setminus \{{\bf x}_{i^\ast}\}\right\} &= & \sum_{i=1}^{i^\ast -1}P_{X^n}({\bf x}_i) < 1 -\varepsilon, \\ \label{eq:4-12}
\Pr\left\{X^n \in L_n \right\} &= &\sum_{i=1}^{i^\ast}P_{X^n}({\bf x}_i) \ge 1 -\varepsilon.
\end{IEEEeqnarray}

We then evaluate the probability 
\begin{IEEEeqnarray}{rCl} \label{eq:4-11-2} 
\Pr\left\{X^n \in  J_n  \cap L_n\right\}& =  &\Pr\left\{X^n \in  J_n \right\} - \Pr\left\{X^n \!\in\!  J_n \cap (L_n)^c\right\}.
\end{IEEEeqnarray}
By the definition of $J_n$, for all $n$ satisfying $\frac{1}{n}\log\frac{1}{P_{X^n}({\bf x}_{i^\ast})} < R$, it holds that
\begin{align} \label{eq:4-13-1}
\Pr\left\{X^n \in  J_n \cap L_n\right\} = 0.
\end{align}
On the other hand, for all $n$ satisfying $\frac{1}{n}\log\frac{1}{P_{X^n}({\bf x}_{i^\ast})} \ge R$, 
\begin{align} \label{eq:4-14}
P_{X^n}({\bf x}_{i^\ast}) \le e^{-nR} 
\end{align}
and
$
(L_n)^c \subseteq J_n
$
hold.
Hence, we have
\begin{align} \label{eq:4-13-2}
\Pr\left\{X^n \!\in\! J_n \cap (L_n)^c\right\}
 & \!=\! \Pr\left\{X^n \in (L_n)^c\right\} \nonumber \\
& \!=\! \Pr\left\{X^n \!\in\! (L_n \!\setminus \{{\bf x}_{i^\ast} \})^c\right\} \!-\! P_{X^n}({\bf x}_{i^\ast}) \nonumber \\
& \!\ge\! \varepsilon - e^{-nR},
\end{align}
where the last inequality is due to (\ref{eq:4-11}) and (\ref{eq:4-14}).
Thus, from (\ref{eq:4-11-2}), (\ref{eq:4-13-1}), and (\ref{eq:4-13-2}) we have 
\begin{IEEEeqnarray}{rCl}
\Pr\left\{X^n \in  J_n \cap L_n\right\} 
& \le & \max\left\{0, \Pr\left\{X^n \in  J_n \right\} \!-\! \left(  \varepsilon \!-\! e^{-nR} \right)  \right\}
\end{IEEEeqnarray}
for all $n$.
Taking $\limsup_{n \to \infty}$ on both sides yields
\begin{IEEEeqnarray}{rCl}  
\lefteqn{\limsup_{n \to \infty}\Pr\left\{X^n \in  J_n \cap L_n\right\} } \nonumber \\
& \le & \max\left\{0, \limsup_{n \to \infty}\Pr\left\{X^n \in  J_n \right\} - \varepsilon \right\}  \le  \delta,
\end{IEEEeqnarray}
where the last inequality is due to (\ref{eq:4-10}).

Since $L_n$ satisfies (\ref{eq:4-12}) we obtain
\begin{align} 
\limsup_{n \to \infty}\inf_{\substack{A_n \subset {\cal X}^n:\\ \Pr\{X^n \in A_n\} \ge 1- \varepsilon}} \Pr\left\{X^n \in  A_n \cap J_n\right\} 
& \le \delta.
\end{align}
This implies $R\ge \tilde{H}_{\varepsilon,\delta}({\bf X})$.
\end{IEEEproof}
\section{Second-order $(\varepsilon,\delta,R)$-optimum threshold}
In this section, we establish the general formula of the second-order $(\varepsilon,\delta,R)$-optimum threshold.
We define the quantity ${G}_{\varepsilon,\delta}(R|{\bf X})$ as
\begin{IEEEeqnarray}{rCl}
\lefteqn{{G}_{\varepsilon,\delta}(R|{\bf X})} \nonumber \\
&  := & \inf\left\{  L \left| \lim_{\nu \downarrow 0}\limsup_{n \to \infty}\inf_{\substack{A_n \subset {\cal X}^n:\\ \Pr\{X^n \in A_n\} \ge 1- \varepsilon - \nu}} \right.  \Pr\left\{ \!- \frac{1}{n}\log \frac{P_{X^n}(X^n)}{\Pr\{X^n \!\in\! A_n\}} \!\ge\! R + \frac{L}{\sqrt{n}}, X^n \!\in\! A_n   \right\} \le \delta  \right\}. \nonumber \\
\end{IEEEeqnarray}
\begin{theorem}[Second-order $(\varepsilon,\delta,R)$-optimum threshold] \label{theo:general2}
For any $\varepsilon, \delta \!\in\! [0,1)$ satisfying $\varepsilon + \delta < 1$, it holds that
\begin{IEEEeqnarray*}{rCl}
L(\varepsilon,\delta,R|{\bf X}) & = {G}_{\varepsilon,\delta}(R|{\bf X}).
\end{IEEEeqnarray*}
\end{theorem}
\begin{IEEEproof}
The proof of the theorem proceeds in parallel with that of Theorem \ref{theo:general1}.
\end{IEEEproof}

As in the \textit{first-order} case (cf. Remark \ref{remark:4-1}), 
${G}_{\varepsilon,\delta}(R|{\bf X})$ can also be expressed as in another way.
Let us define an information-spectrum quantity:
\begin{IEEEeqnarray*}{rCl}
\overline{H}_{\gamma}(R|{\bf X})
& \!:=\! & \inf\left\{ L \left| \limsup_{n \to \infty} \Pr \left\{ \frac{1}{n}\log \frac{1}{P_{X^n}(X^n)} \!>\!R \!+\! \frac{L}{\sqrt{n}}  \right\} \!\le\! \gamma \right. \right\}.
\end{IEEEeqnarray*}
\begin{theorem} \label{theo:another_ex2}
For any $\varepsilon, \delta \in [0,1)$ satisfying $\varepsilon + \delta < 1$, it holds that
\[
{G}_{\varepsilon,\delta}(R|{\bf X}) = \overline{H}_{\varepsilon+\delta}(R|{\bf X}) .
\] 
\end{theorem}
\begin{IEEEproof} The proof is similar to that of Theorem \ref{theo:another_ex}.
\end{IEEEproof}

The \textit{second-order} $(\varepsilon,R)$-optimum coding rate in the fixed-length coding is characterized by $\overline{H}_{\varepsilon}(R|{\bf X})$ \cite{Hayashi}.
This means that in order to compute ${G}_{\varepsilon,\delta}(R|{\bf X})$ for some specified sources such as i.i.d. sources and Markov sources, we can use the similar technique as the one in \cite{Hayashi} (see, Section VII).
%
%
%
%
%
%
\section{Optimistic optimum thresholds}
\subsection{General Formulas}
So far, we have considered the first- and second-order optimum thresholds.
In this section, we establish the coding theorems in the optimistic sense, which will turn out to reveal another relationship with the fixed-length source coding.
The notion of the optimistic coding has first been introduced by Vembu, Verd\'u and Steinberg \cite{VVS} and several researchers have developed the optimistic coding scenario in other information theoretic problems \cite{Hayashi,KOGA2014,YN2017_7}.
We also develop the notion of the optimistic coding to the $\varepsilon$-variable-length coding problem.
\begin{definition}
Rate $R$ is said to be optimistically $(\varepsilon,\delta)$-achievable $(\varepsilon, \delta \in [0,1))$, if there exists a sequence of variable-length code $(\varphi_n,\psi_n)$ such that for any $\gamma >0$
\begin{IEEEeqnarray}{rCl} \label{eq:def1-1-2}
\varepsilon_{n_i} \le \varepsilon + \gamma, \quad 
\delta_{n_i} \left(n_i R\right) \le \delta + \gamma
\end{IEEEeqnarray}
hold for some subsequence $n_1 < n_2 < \cdots$.
\end{definition}
\begin{definition}[Optimistic first-order $(\varepsilon,\delta)$-optimum threshold]
\begin{equation*}
{R}^\ast(\varepsilon,\delta|{\bf X}) := \inf\left\{ R | R \mbox{ is optimistically $(\varepsilon,\delta)$-achievable}  \right\}.
\end{equation*}
\end{definition}

\begin{definition}
Rate $L$ is said to be optimistically $(\varepsilon,\delta, R)$-achievable $(\varepsilon, \delta \in [0,1), R\ge 0)$, if there exists a sequence of variable-length code $(\varphi_n,\psi_n)$ such that for any $\gamma >0$
\begin{IEEEeqnarray}{rCl} \label{eq:def2-1-2}
\varepsilon_{n_i} \le \varepsilon + \gamma, \quad 
\delta_{n_i} \left({n_i}R + {\sqrt{n_i}}{L}\right) \le \delta + \gamma
\end{IEEEeqnarray}
hold for some subsequence $n_1 < n_2 < \cdots$.
\end{definition}
\begin{definition}[Optimistic second-order $(\varepsilon,\delta,R)$-optimum threshold]
\begin{equation}
{L}^\ast(\varepsilon,\delta, R|{\bf X}) := \inf\{L | L \mbox{ is  optimistically }(\varepsilon,\delta,R)\mbox{-achievable} \}.
\end{equation}
\end{definition}

\begin{remark}
One may consider that we can define different quantities as follows:
\begin{definition}
Rate $R$ is said to be type I $(\varepsilon,\delta)$-achievable $(\varepsilon, \delta \in [0,1))$, if there exists a sequence of variable-length code $(\varphi_n,\psi_n)$ such that
\begin{equation} 
\limsup_{n \to \infty} \varepsilon_n \le \varepsilon, \quad
\liminf_{n \to \infty} \delta_n(nR) \le \delta.
\end{equation}
\end{definition}
\begin{definition} 
\begin{equation}
{R}^\dagger(\varepsilon,\delta|{\bf X}) := \inf\left\{ R | R \mbox{ is type I $(\varepsilon,\delta)$-achievable}  \right\}.
\end{equation}
\end{definition}
\begin{definition}
Rate $R$ is said to be type II $(\varepsilon,\delta)$-achievable $(\varepsilon, \delta \in [0,1))$, if there exists a sequence of variable-length code $(\varphi_n,\psi_n)$ such that
\begin{equation} 
\liminf_{n \to \infty} \varepsilon_n \le \varepsilon, \quad
\limsup_{n \to \infty} \delta_n(nR) \le \delta.
\end{equation}
\end{definition}
\begin{definition} 
\begin{equation}
{R}^\ddagger(\varepsilon,\delta|{\bf X}) := \inf\left\{ R | R \mbox{ is type II $(\varepsilon,\delta)$-achievable}  \right\}.
\end{equation}
\end{definition}
Then, from the definitions it is not difficult to check that
\begin{equation}
{R}^\dagger(\varepsilon,\delta|{\bf X})  = {R}^\ddagger(\varepsilon,\delta|{\bf X}) = {R}^\ast(\varepsilon,\delta|{\bf X}). 
\end{equation}
The similar relationship also holds for the optimistic \textit{second-order} optimum thresholds. \IEEEQED
\end{remark}
In order to characterize ${R}^\ast(\varepsilon,\delta|{\bf X})$ and ${L}^\ast(\varepsilon,\delta,R|{\bf X})$, we define the information theoretic quantities as follows.
\begin{multline}
{G}^\ast_{\varepsilon,\delta}({\bf X})  := \inf\left\{  R \left| \lim_{\nu \downarrow 0}\liminf_{n \to \infty}\inf_{\substack{A_n \subset {\cal X}^n:\\ \Pr\{X^n \in A_n\} \ge 1- \varepsilon - \nu}} \right. \Pr\left\{ \!- \frac{1}{n}\log \frac{P_{X^n}(X^n)}{\Pr\{X^n \!\in\! A_n\}} \!\ge\! R, X^n \!\in\! A_n   \right\} \le \delta  \right\},
\end{multline}
\begin{IEEEeqnarray}{rCl}
\lefteqn{{G}^\ast_{\varepsilon,\delta}(R|{\bf X})} \nonumber \\
&  := & \inf\left\{  L \left| \lim_{\nu \downarrow 0}\liminf_{n \to \infty}\inf_{\substack{A_n \subset {\cal X}^n:\\ \Pr\{X^n \in A_n\} \ge 1- \varepsilon - \nu}} \right. \Pr\left\{ \!- \frac{1}{n}\log \frac{P_{X^n}(X^n)}{\Pr\{X^n \!\in\! A_n\}} \!\ge\! R + \frac{{L}}{\sqrt{n}}, X^n \!\in\! A_n   \right\} \le \delta  \right\}. \nonumber \\
\end{IEEEeqnarray}
Then, we have the following theorem which provides the \textit{general formulas} of the first- and second-order optimum thresholds in the optimistic coding problem.
\begin{theorem} \label{theo:general1_o}
For any $\varepsilon, \delta \in [0,1)$ satisfying $\varepsilon + \delta < 1$, it holds that
\begin{align} \label{eq:general1_o}
{R}^\ast(\varepsilon,\delta|{\bf X}) & = {G}^\ast_{\varepsilon,\delta}({\bf X}), \\ \label{eq:general2_o}
{L}^\ast(\varepsilon,\delta,R|{\bf X})  &= {G}^\ast_{\varepsilon,\delta}(R|{\bf X}).
\end{align}
\end{theorem}
\begin{IEEEproof}
We here only show (\ref{eq:general2_o}). The proof of (\ref{eq:general1_o}) is similar to that of (\ref{eq:general2_o}).

(Direct Part:)
Setting $
L_0  = {G}^\ast_{\varepsilon,\delta}(R|{\bf X}),
$
 we show that $L=L_0 + 2\gamma$ is optimistically $(\varepsilon,\delta,R)$-achievable for any $\gamma >0$.

To do so, we arbitrarily fix $\nu >0$ and consider a subsequence $\{n_1 < n_2 < \dots \}$ satisfying
\begin{IEEEeqnarray}{rCl}  \label{eq:d-0-0}
\lim_{i \to \infty}\inf_{\substack{A_{n_i} \subset {\cal X}^{n_i}:\\ \Pr\{X^{n_i} \in A_{n_i}\} \ge 1- \varepsilon - \nu}}
 \Pr\left\{ - \frac{1}{{n_i}}\log \frac{P_{X^{n_i}}(X^{n_i})}{\Pr\{X^{n_i} \in A_{n_i}\}} \ge a + \frac{L_0 + \gamma}{\sqrt{n}}, X^{n_i} \in A_{n_i}  \right\} \le \delta.
\end{IEEEeqnarray}
Moreover, let $\lambda_1 > \lambda_2 > \dots \to 0$ be an arbitrary monotone decreasing sequence and choose
$A_n \subseteq {\cal X}^n$ such that 
\begin{IEEEeqnarray}{rCl}
\Pr\left\{  X^{n_i} \in A_{n_i} \right\} \ge 1 - \varepsilon - \nu
\end{IEEEeqnarray}
and
\begin{IEEEeqnarray}{rCl} \label{eq:d-0}
\lefteqn{ \Pr \left\{  \! -\!\frac{1}{n_i} \log \frac{P_{X^{n_i}}({X^{n_i}})}{ \Pr \left\{ X^{n_i} \!\in\! A_{n_i} \right  \}} \ge a \!+\! \frac{L_0 \!+\! \gamma}{\sqrt{n_i}}, \ X^{n_i} \in A_{n_i
} \right\}} \nonumber  \\
 & \le  \inf_{\substack{A_{n_i} \subset {\cal X}^{n_i}:\\ \Pr\{X^{n_i} \in A_{n_i}\} \ge 1- \varepsilon - \nu}} \Pr\left\{ \! -\!\frac{1}{n_i} \log \frac{P_{X^{n_i}}({X^{n_i}})}{ \Pr \left\{ X^{n_i} \!\in\! A_{n_i} \right  \}} \ge a \!+\! \frac{L_0 \!+\! \gamma}{\sqrt{n_i}}, \  X^{n_i} \!\in\! A_{n_i} \right\} \!+\!\lambda_{n_i}
\end{IEEEeqnarray}
hold.

Then, Theorem \ref{theo:direct} with $a_n = K^{-\sqrt{n} \gamma}$，$\eta_n= na + \sqrt{n}L=na + \sqrt{n}(L_0 + 2\gamma)$ guarantees that there exists a variable-length code $(\varphi_n,\psi_n)$ such that
\begin{equation}
\varepsilon_{n_i} \le \varepsilon + \nu,
\end{equation}
and
\begin{align*}
\lefteqn{ \Pr \left\{ \frac{1}{n_i} l(\varphi_{n_i}(X^{n_i})) > a + \frac{L_0 + 2\gamma}{\sqrt{n_i}} \right\}} \\
& \!\le\! \Pr\left\{  -\frac{1}{n_i} \log \frac{P_{X^{n_i}}({X^{n_i}})}{ \Pr \left\{ X^{n_i} \in A_{n_i} \right  \}} \ge a+ \frac{L_0 \!+\! \gamma}{\sqrt{n_i}} , \  X^{n_i} \in A_{n_i} \right\}  + K^{-\sqrt{n_i}\gamma +1}
\end{align*}
hold for subsequence $\{n_i\}_{i=1}^\infty$ .

Since $A_n$ satisfies (\ref{eq:d-0}), it holds that 
\begin{equation}  \label{eq:d-1}
\varepsilon_{n_i} \le \varepsilon + \nu,
\end{equation}
and
\begin{align} \label{eq:d-2-2}
\lefteqn{ \Pr \left\{ \frac{1}{n_i} l(\varphi_n(X^{n_i})) >a + \frac{L_0 + 2\gamma}{\sqrt{n_i}} \right\}} \nonumber  \\
 & \le  \inf_{\substack{A_{n_i} \subset {\cal X}^{n_i}:\\ \Pr\{X^{n_i} \in A_{n_i}\} \ge 1- \varepsilon - \nu}}  \Pr\left\{ \! -\!\frac{1}{{n_i}} \log \frac{P_{X^{n_i}}({X^{n_i}})}{ \Pr \left\{ X^{n_i} \!\in\! A_{n_i} \right  \}} \ge a \!+\! \frac{L_0 \!+\! \gamma}{\sqrt{n_i}}, \  X^{n_i} \!\in\! A_{n_i} \right\}  \nonumber \\
&     \quad + K^{-\sqrt{n_i}\gamma +1} +  \lambda_{n_i} \nonumber \\
& \le \delta + \nu, \quad (\forall i > i_0),
\end{align}
where the last inequality is due to (\ref{eq:d-0-0}) and the fact that $\lambda_{n_i} \to 0\ (i \to \infty)$.
Hence, from (\ref{eq:d-1}) and (\ref{eq:d-2-2}) the direct part has been proved.

(Converse Part:)
We assume that $L$ is optimistically $(\varepsilon,\delta,R)$-achievable.
This means that for any $\nu >0$ there exists a variable-length code $(\varphi_n, \psi_n)$ such that  
\begin{align}  \label{eq:c-1-2}
\Pr \left\{ X^{n_i} \neq \psi_{n_i}(\varphi_n(X^{n_i})) \right\} &\le \varepsilon + \nu, \\ \label{eq:c-2-2}
\Pr \left\{ \frac{1}{n_i} l(\varphi_{n_i}(X^{n_i})) > R + \frac{L}{\sqrt{n_i}} \right\} &\le \delta + \nu,
\end{align}
holds for a subsequence $\{n_i \}_{i=1}^\infty$.
We fix this subsequence $\{n_i \}_{i=1}^\infty$ and the code $(\varphi_n, \psi_n)$.

Let $\gamma >0$ be any fixed constant. Then, by using Theorem \ref{theo:converse} with $a_n = K^{-\sqrt{n} \gamma}$, $\eta_n= nR + \sqrt{n}L$, we obtain
\begin{align} \label{eq:d-1-1} 
 \lefteqn{\Pr \left\{ \frac{1}{n_i}l(\varphi_{n_i}(X^{n_i})) > R + \frac{L}{\sqrt{n_i}} \right\}} \nonumber \\
 & \!\ge\!  \Pr\left\{  \frac{P_{X^{n_i}}({X^{n_i}})}{ \Pr \left\{ X^{n_i} \!\in\! D_{n_i} \right  \}} \le K^{-{n_i} R - \sqrt{n_i}(L + \gamma)}, \  X^{n_i} \!\in\! D_{n_i} \right\} -\! K^{-\sqrt{n_i}\gamma +1} \nonumber \\
 & \ge  \inf_{\substack{A_{n_i} \subset {\cal X}^{n_i}:\\ \Pr\{X^{n_i} \in A_{n_i}\} \ge 1- \varepsilon - \nu}} \Pr\left\{ -\frac{1}{n_i} \log \frac{P_{X^{n_i}}({X^{n_i}})}{ \Pr \left\{ X^{n_i} \in A_{n_i} \right  \}} \ge R + \frac{L \!+\! \gamma}{\sqrt{n_i}}, \  X^{n_i} \in A_{n_i} \right\},
\end{align}
since
$
D_n := \left\{ {\bf x} \in {\cal X}^n \left| {\bf x} = \psi_n(\varphi_n({\bf x})) \right. \right\}
$
satisfies 
$ \Pr\left\{ X^{n_i} \in D_{n_i} \right\} \ge 1 - \varepsilon - \nu $ from (\ref{eq:c-1-2}).

Then from (\ref{eq:c-2-2}) and (\ref{eq:d-1-1}) we have for $i=1,2,\dots,$
\begin{IEEEeqnarray}{rCl} 
\delta + \nu  \ge   \inf_{\substack{A_{n_i} \subset {\cal X}^{n_i}:\\ \Pr\{X^{n_i} \in A_{n_i}\} \ge 1- \varepsilon - \nu}} \Pr\left\{ \!-\frac{1}{n_i} \log \frac{P_{X^{n_i}}({X^{n_i}})}{ \Pr \left\{ X^{n_i} \!\in\! A_{n_i} \right  \}} \ge R + \frac{L+ \gamma}{\sqrt{n_i}}, \  X^{n_i} \!\in\! A_{n_i} \right\}
 \end{IEEEeqnarray}
from which it holds that
\begin{IEEEeqnarray}{rCl}
\lefteqn{L + \gamma } \nonumber \\
& \ge &  \inf\left\{  L \left| \lim_{\nu \downarrow 0}\liminf_{n \to \infty}\inf_{\substack{A_n \subset {\cal X}^n:\\ \Pr\{X^n \in A_n\} \ge 1- \varepsilon - \nu}} \right. 
\Pr\left\{ - \frac{1}{n}\log \frac{P_{X^n}(X^n)}{\Pr\{X^n \in A_n\}} \ge R + \frac{L}{\sqrt{n}}, X^n \in A_n   \right\} \le \delta  \right\}. \nonumber \\
\end{IEEEeqnarray}
This means that the converse part holds.
\end{IEEEproof}
\subsection{Relations to the fixed-length source coding problem}
In this subsection we reveal the relationship between the optimistically optimum thresholds with the optimistically optimum fixed-length source coding rates.

Let $\varphi^f_n: {\cal X}^n\to{\cal M}_n:=\{1,2,\dots,M_n\}$ and $\psi^f_n: {\cal M}_n \to {\cal X}^n$ be a fixed-length encoder and a decoder, respectively.
Then, the error probability is defined by
$
\varepsilon_n:= \Pr \left\{ X^n \neq \psi_n^f(\varphi_n^f(X^n))\right\}.
$
We call such a code $(n, M_n,\varepsilon_n)$ code.
\begin{definition} 
Rate $R$ is said to be optimistically $\varepsilon$-achievable if there exists a sequence of fixed-length code $(n,M_n,\varepsilon_n)$ such that for any $\gamma >0$
\begin{equation}
\varepsilon_{n_i} \leq \varepsilon + \gamma , \quad \frac{1}{{n_i}} \log M_{n_i} \leq R + \gamma
\end{equation}
hold for some subsequence $n_1 < n_2 < \cdots$.
\end{definition}
\begin{definition}
\begin{equation}
R^f(\varepsilon |{\bf X}) :=  \inf \left\{ R \left|R \mbox{ is optimistically $\varepsilon$-achievable} \right. \right\}.
\end{equation}
\end{definition}
\begin{definition} 
Rate $L$ is said to be optimistically $(\varepsilon,R)$-achievable if there exists a sequence of fixed-length code $(n,M_n,\varepsilon_n)$ such that for any $\gamma >0$
\begin{equation}
\varepsilon_{n_i} \leq \varepsilon + \gamma , \quad \frac{1}{\sqrt{n_i}} \log \frac{{M_{n_i}}}{K^{{n_i}R}} \leq L + \gamma
\end{equation}
hold for some subsequence $n_1 < n_2 < \cdots$.
\end{definition}
\begin{definition}
\begin{equation}
L^f(\varepsilon,R |{\bf X}) :=  \inf \left\{ L \left|L \mbox{ is optimistically $(\varepsilon,R)$-achievable} \right. \right\}.
\end{equation}
\end{definition}
General formulas of the above quantities have been given by Chen and Alajaji \cite{Chen1999} and Hayashi \cite{Hayashi} as follows.
\begin{theorem}[Chen and Alajaji\cite{Chen1999}, Hayashi\cite{Hayashi}]
\begin{IEEEeqnarray}{rCl}
R^f(\varepsilon |{\bf X}) = \overline{H}^\ast_{\varepsilon}({\bf X}), \quad 
L^f(\varepsilon,R|{\bf X}) = \overline{H}^\ast_{\varepsilon}(R|{\bf X}),
\end{IEEEeqnarray}
where
\begin{align}
\overline{H}^\ast_{\gamma}({\bf X}) & := \inf\left\{ R \left| \liminf_{n \to \infty} \Pr \left\{ \frac{1}{n}\log \frac{1}{P_{X^n}(X^n)} > R  \right\} \le \gamma \right. \right\}, \\
\overline{H}^\ast_{\gamma}(R|{\bf X}) & := \inf\left\{ L \left| \liminf_{n \to \infty} \Pr \left\{ \frac{1}{n}\log \frac{1}{P_{X^n}(X^n)} >R + \frac{L}{\sqrt{n}}  \right\} \le \gamma \right. \right\}.
\end{align}
\end{theorem}
Then, we obtain the following theorem which is analogous to Theorem \ref{theo:another_ex}.
\begin{theorem}
For any $\varepsilon,\delta \in [0,1)$ satisfying $\varepsilon +\delta <1$, it holds that
\begin{align}
G^\ast_{\varepsilon,\delta}({\bf X}) = \overline{H}^\ast_{\varepsilon+\delta}({\bf X}), \quad G^\ast_{\varepsilon,\delta}(R|{\bf X}) = \overline{H}^\ast_{\varepsilon+\delta}(R|{\bf X}).
\end{align}
\end{theorem}
\begin{IEEEproof}
The proof proceeds in parallel with the proof of Theorem \ref{theo:another_ex}.
\end{IEEEproof}
From the above theorem and Theorem \ref{theo:general1_o}, 
we immediately obtain the following corollary.
\begin{corollary} \label{coro:4}
For any $\varepsilon,\delta \in [0,1)$ satisfying $\varepsilon +\delta <1$ it holds that
\begin{align}
R^\ast(\varepsilon,\delta|{\bf X}) &  = R^f(\varepsilon + \delta |{\bf X}) = \overline{H}^\ast_{\varepsilon+\delta}({\bf X}) , \\
 L^\ast(\varepsilon,\delta,R|{\bf X}) & = L^f(\varepsilon + \delta,R |{\bf X})  = \overline{H}^\ast_{\varepsilon+\delta}(R|{\bf X}).
\end{align} \IEEEQED
\end{corollary}
This corollary shows that the optimistically $(\varepsilon,\delta)$-optimum thresholds coincide with the optimistically $(\varepsilon+\delta)$-optimum fixed-length coding rate for any $\varepsilon,\delta \in [0,1)$.
Hence, we can compute $R^\ast(\varepsilon,\delta|{\bf X})$ and  $L^\ast(\varepsilon,\delta,R|{\bf X})$ by using the results of the optimistic $\gamma$-fixed-length coding problem such as \cite{Chen1999,Hayashi} (see, Section VII).
%
%
%
%
%
%
\section{Application to Stationary memoryless sources}
In this section we compute the optimum overflow thresholds for the stationary memoryless source with generic distribution $X$ by using \textit{general formulas} obtained in the preceding sections.
\subsection{First-Order $(\varepsilon,\delta)$-Optimum Thresholds}
For a stationary memoryless source $X$, the following theorem is well-known.
\begin{theorem}[Chen and Alajaji \cite{Chen1999}, Steinberg and Verd\'{u} \cite{Steinberg}] \label{theo:iid1}
For any $\gamma \in [0,1) $, it holds that
\begin{align}
\overline{H}_{\gamma}({\bf X}) = \overline{H}^\ast_{\gamma}({\bf X}) = H(X), 
\end{align}
where $H(X)$ denotes the entropy of the source $X$. 
\end{theorem}
From Theorems \ref{theo:general1}, \ref{theo:another_ex}, and \ref{theo:iid1}, and Corollary \ref{coro:4} we immediately have
\begin{theorem}
For any $ \varepsilon, \delta  \in [0,1)$ satisfying $\varepsilon + \delta <1$ it holds that
\begin{IEEEeqnarray}{rCl}
R(\varepsilon,\delta|{\bf X}) = R^\ast(\varepsilon,\delta|{\bf X}) = H(X) \quad (0 \le \varepsilon, \delta  <1).
\end{IEEEeqnarray}
\end{theorem}
Thus, the $(\varepsilon,\delta)$-optimum overflow thresholds and the optimistically $(\varepsilon,\delta)$-optimum overflow thresholds coincides and equals to the entropy of the source irrespective of $\varepsilon$ and $\delta$.
\subsection{Second-Order $(\varepsilon,\delta,R)$-Optimum Overflow Thresholds}
In the second-order coding rate analysis, it is well-known that:
\begin{theorem}[Hayashi\cite{Hayashi}] \label{theo:iid2}
\begin{align}
\overline{H}_{\gamma}(R|X)  = \overline{H}_{\gamma}^\ast(R|X) = \left\{ \begin{array}{ll} 
-\infty & R > H(X) \\
+ \infty & R < H(X) \\
\sqrt{V_X}\Phi^{-1} \left({\gamma} \right) & R = H(X),
 \end{array}\right.
\end{align}
where
\begin{equation}
\Phi(x) :=  \frac{1}{\sqrt{2\pi}}\int^{\infty}_x e^{-\frac{x^2}{2}} dx
\end{equation}
is the standard Gaussian distribution function and 
\begin{equation}
V_X := \sum_{x \in {\cal X}} P_{X}(x) \left( - \log {P_{X}(x)}- H(X) \right)^2.
\end{equation}
denotes the variance of the self-information called \textit{varentropy} of the source \cite{KV2014}. \IEEEQED
\end{theorem}
As is known from the above theorem, the setting of \textit{first-order} constant $R$ is quite important to analyze the second-order $(\varepsilon,\delta,R)$-optimum overflow thresholds.
In this paper, we consider the following two case:

Case 1) Setting $R$ as the first-order $(\varepsilon,\delta)$-optimum thresholds $R_1$ and the first-order optimistically $(\varepsilon,\delta)$-optimum thresholds $R_1^\ast$:

In this case, from Theorem \ref{theo:general1} and Corollary \ref{coro:4}
\begin{align}
R_1  =  \overline{H}_{\varepsilon+ \delta}({\bf X}), \quad  R_1^\ast = \overline{H}^\ast_{\varepsilon+ \delta}({\bf X})
\end{align}
hold.
Thus, from Theorem \ref{theo:iid1} we set
\begin{align}
R_1 = R_1^\ast = H(X)
\end{align}
for the stationary memoryless source with generic distribution $X$.

Case 2) Setting $R$ as the optimum mean codeword length $R_2$ and optimistically optimum mean codeword length $R_2^\ast$: 

The optimum mean codeword length of the $\varepsilon$-variable-length codes has been first determined by Koga and Yamamoto \cite{KY2005} in the case that $\varepsilon \in [0,1)$, while Han \cite{Han2000_wv, Han} has derived it in the case of $\varepsilon=0$.
Yagi and Nomura\cite{YN2017_7} have determined the optimistically optimum mean codeword length of the $\varepsilon$-variable-length codes.

From results in \cite{KY2005} and \cite{YN2017_7}, we shall set
\begin{IEEEeqnarray}{rCl}
R_2 & =&  H_{[\varepsilon]}({\bf X}):= \lim_{\nu \downarrow 0}\limsup_{n \to \infty} \frac{1}{n} \inf_{\substack{A_n \subset {\cal X}^n:\\ \Pr\{X^n \in A_n\} \ge 1- \varepsilon - \nu}}\sum_{{\bf x} \in A_n} P_{X^n}({\bf x})\log \frac{1}{P_{X^n}({\bf x})}, \\
R_2^\ast & = & \lim_{\nu \downarrow 0}\liminf_{n \to \infty} \frac{1}{n} \inf_{\substack{A_n \subset {\cal X}^n:\\ \Pr\{X^n \in A_n\} \ge 1- \varepsilon - \nu}}\sum_{{\bf x} \in A_n} P_{X^n}({\bf x})\log \frac{1}{P_{X^n}({\bf x})}.
\end{IEEEeqnarray}
In particular, 
\begin{equation}
R_2 = R_2^\ast = (1 - \varepsilon)H(X)
\end{equation}
holds for the stationary memoryless source with generic distribution $X$\cite{KY2005,YN2017_7}．
\begin{remark}
The optimum {\it second-order} mean codeword length $L_{[\varepsilon]}(H_{[\varepsilon]}({\bf X})|{\bf X})$ for the stationary memoryless source $X$ has been determined by Kostina, Polyanskiy and Verd\'{u} \cite{KV2015} as follows.

Assuming that the third absolute moment of $-\log P_X(X)$ is finite, then for any given $\varepsilon \in (0,1)$, it holds that
\begin{equation}
L_{[\varepsilon]}(H_{[\varepsilon]}({\bf X})|{\bf X}) = - \sqrt{\frac{V(X)}{2\pi}}e^{- \frac{ (\Phi^{-1}(\varepsilon) )^2}{2}}.
\end{equation}
\end{remark}
The above result shows an interesting phenomenon in which the optimum {\it second-order} mean codeword length is always negative.

Summarizing up, we set $R$ as $R_1 = H(X)$ and $R_2 = (1-\varepsilon)H(X)$.

Then, we obtain the following theorem.
\begin{theorem} \label{theo:iid3}
For any $\varepsilon,\delta \in [0,1)$ satisfying $\varepsilon+\delta<1$, it holds that
\begin{IEEEeqnarray}{rCl}  \label{eq:5-1}
L(\varepsilon,\delta,R_1|{\bf X}) & = & L^\ast(\varepsilon,\delta,R^\ast_1|{\bf X}) = \sqrt{V_X}\Phi^{-1}\left(\varepsilon + \delta \right), \\ \label{eq:5-2} 
L(\varepsilon,\delta,R_2|{\bf X}) & = &L^\ast(\varepsilon,\delta,R^\ast_2|{\bf X}) = \left\{ \begin{array}{ll}
\sqrt{V_X}\Phi^{-1}\left(\delta\right) & \varepsilon = 0 \\
+ \infty & \varepsilon \neq 0.
\end{array}\right. 
\end{IEEEeqnarray}
\end{theorem}
\begin{IEEEproof}
From Theorems \ref{theo:general2}, \ref{theo:another_ex2}, and \ref{theo:iid2} we obtain (\ref{eq:5-1}) as well as  (\ref{eq:5-2}) in the case of $\varepsilon=0$. 
On the other hand, when we consider the case of $L(\varepsilon,\delta,R_2|{\bf X})$ with $\varepsilon > 0$, it holds that $R_2=(1 - \varepsilon)H(X) < {H}({X})$.
Thus, from Theorems \ref{theo:general2} and \ref{theo:another_ex2}, and the definition of $\overline{H}_{\varepsilon+\delta}(R_2|{\bf X})$ it holds that
\begin{IEEEeqnarray}{rCl}
L(\varepsilon,\delta,R_2|{\bf X}) & = & \overline{H}_{\varepsilon+\delta}(R_2|{\bf X}) \nonumber \\
&  = &\inf\left\{ L \left|  \limsup_{n \to \infty} \Pr\left\{ \frac{1}{n} \log \frac{1}{P_{X^n}(X^n)} > R_2 + \frac{L}{\sqrt{n}}\right\} \le \varepsilon+ \delta \right.\right\} \nonumber \\
& = & \inf\left\{ L \left|  \limsup_{n \to \infty} \Pr\left\{ \frac{1}{n} \log \frac{1}{P_{X^n}(X^n)} > (1 - \varepsilon)H(X) + \frac{L}{\sqrt{n}}\right\} \le \varepsilon+ \delta \right.\right\}.
\end{IEEEeqnarray}
Here, from the law of large numbers this case necessarily yields that 
\begin{IEEEeqnarray}{rCl}
\lim_{n \to \infty}\Pr\left\{ \frac{1}{n} \log \frac{1}{P_{X^n}(X^n)} > (1 - \varepsilon)H(X) + \frac{L}{\sqrt{n}}\right\} = 1,
\end{IEEEeqnarray}
for any constant $L < \infty$.
Hence, in this case we set formally as $L(\varepsilon,\delta,R_2|{\bf X}) = +\infty$.
We similarly obtain the result in the case of $L^\ast(\varepsilon,\delta,R_2^\ast|{\bf X}) = \overline{H}^\ast_{\varepsilon+\delta}(R_2^\ast|{\bf X})$.
\end{IEEEproof}

The above theorem shows that if we set $R$ as $R= R_2$ (the optimum mean codeword length of $\varepsilon$-variable-length codes), the second-order $(\varepsilon,\delta,R)$-optimum thresholds is always infinity as long as $\varepsilon > 0$.
This means that the error probability or the overflow probability cannot be less than or equal to the desired value irrespective with the second-order threshold $L$.
Moreover, from the similar argument to the proof of Theorem \ref{theo:iid3}, we observe that $L(\varepsilon,\delta,R|{\bf X})=L^\ast(\varepsilon,\delta,R|{\bf X})= \infty$ for $R < H(X)$, and $L(\varepsilon,\delta,R|{\bf X})=L^\ast(\varepsilon,\delta,R|{\bf X})= - \infty$ for $R> H(X)$.  
Hence, in order to analyze the \textit{second-order} optimum threshold $L$ in the $\varepsilon$-variable-length coding, the \textit{first-order} rate $R$ should be set as the first-order optimum threshold: $R_1 = H(X)$.
\section{Concluding Remarks}
We have so far considered the \textit{first}- and \textit{second-order} achievability to evaluate several types of  optimum overflow thresholds in the $\varepsilon$-variable-length coding problem. 
As shown in the proofs of this paper, the information spectrum approach is substantial in analyses. In particular, Theorems \ref{theo:direct} and \ref{theo:converse} enable us to analyze the all types of the optimum overflow thresholds by the unified approach.

We have also clarified that the relationship between the $\varepsilon$-variable-length coding and the $\gamma$-fixed-length coding.
In the $\gamma$-fixed-length coding problem, the $\gamma$-optimum coding rate has already been derived for several tractable sources such as stationary memoryless sources, Markov sources and mixed sources \cite{Hayashi,NH2011}.
We can use these results to compute the $(\varepsilon,\delta)$-optimum thresholds as well as the $(\varepsilon,\delta,R)$-optimum thresholds in the $\varepsilon$-variable-length coding.
Actually, in this paper we compute the optimum thresholds in the $\varepsilon$-variable-length coding for the stationary memoryless source by using the previous results for the $\gamma$-fixed-length coding.
\section*{Acknowledgement}
This work was supported in part by JSPS KAKENHI No. JP16K06340 and No. JP18K04150.




\ifCLASSOPTIONcaptionsoff
  \newpage
\fi

\end{document}